# Re: What's Up Johnny?

## Covert Content Attacks on Email End-to-End Encryption

Jens Müller[1], Marcus Brinkmann[1], Damian Poddebniak[2], Sebastian Schinzel[2], and Jörg Schwenk[1]

[1] Ruhr University Bochum, Germany
{jens.a.mueller,marcus.brinkmann,joerg.schwenk}@rub.de
[2] Münster University of Applied Sciences, Germany
{damian.poddebniak,schinzel}@fh-muenster.de

**Abstract.** We show practical attacks against OpenPGP and S/MIME encryption and digital signatures in the context of email. Instead of targeting the underlying cryptographic primitives, our attacks abuse legitimate features of the MIME standard and HTML, as supported by email clients, to deceive the user regarding the actual message content. We demonstrate how the attacker can unknowingly abuse the user as a decryption oracle by replying to an unsuspicious looking email. Using this technique, the plaintext of hundreds of encrypted emails can be leaked at once. Furthermore, we show how users could be tricked into signing arbitrary text by replying to emails containing CSS conditional rules. An evaluation shows that 17 out of 19 OpenPGP-capable email clients, as well as 21 out of 22 clients supporting S/MIME, are vulnerable to at least one attack. We provide different countermeasures and discuss their advantages and disadvantages.

## 1 Introduction

Email was designed as a plaintext protocol, which allows eavesdroppers to read or modify the communication on the channel. While it is common today that traffic between mailservers is TLS encrypted,[3] transport encryption is not sufficient to protect against strong attackers, such as a man-in-the-middle (MitM) within the infrastructure (e.g., a dishonest mail server operator), or an attacker who gains access to leaked user emails. OpenPGP [2] and S/MIME [9] are the two major standards used in such scenarios and provide end-to-end cryptographic protection. Both standards are designed to guarantee confidentiality, integrity, and authenticity of messages, even in hostile environments such as a compromised or untrustworthy mail server by encrypting and digitally signing emails.

---

[3] According to Google's transparency report, 88% of the email traffic was TLS encrypted in the fourth quarter of 2018: https://transparencyreport.google.com/safer-email/

**Research Question.** Both standards are based on asymmetric encryption; only the user has access to the private key and, therefore, can decrypt messages encrypted with the public key or sign messages. However, email usage involves interaction with multiple communication partners, including potentially dishonest parties. Example: a mail server operator, Eve, who is in possession of the ciphertext messages sent from Alice to Bob can simply re-send the encrypted message from her address and have Bob decrypt it.[4] If Bob simply replied to Eve while quoting the original message, he would leak the plaintext of his communication with Alice. Such message takeover attacks under a new identity are well-known issues in email end-to-end encryption (see [6, 7]). However, they are generally considered an acceptable risk because it is assumed that given the context of the message (e.g., *"Hi Bob, [...] Yours, Alice"*) Bob can tell that this message is not originally from Eve and could easily discover the deception.

Therefore, the research question arises: *Is it possible to hide the original text to trick a user into unintentionally acting as a decryption oracle?* A schematic illustration of such an attack is given in Figure 1.

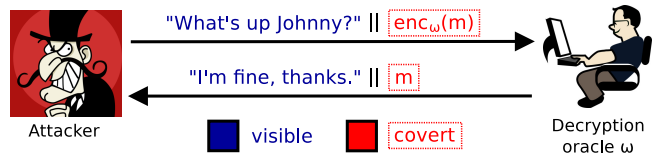

Fig. 1: Covert content attacks against email encryption.

**Contributions.** In this work, we show simple, yet practical, attacks against email encryption and digital signatures, and discuss the countermeasures. We demonstrate how an attacker can wrap ciphertext into a specially crafted email which looks benign but leaks the plaintext of hundreds of encrypted emails at once if replied to. Furthermore, we show how to turn the victim into a signing oracle by having him sign quoted covert content. The attacker can put this content into a different context based on CSS conditional rules, resulting in arbitrary text to be displayed as correctly signed by the victim. Our evaluation shows that 17 out of 19 OpenPGP capable email clients, as well as 21 out of 22 clients supporting S/MIME are vulnerable to at least one attack. Our attacks raise concerns about the overall security of encryption and digital signatures in the context of email, even though the security guarantees of the cryptography behind them remains untouched.

**Responsible Disclosure.** We reported our attacks to the affected vendors and proposed appropriate countermeasures. Our findings regarding email end-to-end encryption resulted in CVE-2019-10731 to CVE-2019-10741. Our attacks on digital signatures are documented as CVE-2019-10726 to CVE-2019-10730.

---

[4] Note that digital signatures do not prevent this attack because Eve can strip them and re-sign the message under her identity as discussed in section 8.1 of this paper.

## 2    Background

In this section, we provide the fundamentals and the historical context of the OpenPGP and S/MIME encryption schemes, as well as MIME and HTML email.

### 2.1    OpenPGP

Pretty Good Privacy (PGP) was invented in 1991 by Phil Zimmermann and played a major political role in the 'crypto wars' of the mid-1990s. Until today, it has a high reputation among activists, journalists, and privacy enthusiasts. PGP was standardized as OpenPGP in RFC4880 which comes in two flavors: For PGP/Inline, the plaintext in the email body is simply replaced by its encrypted counterpart. This is done separately for each body part (or attachment) in case of multipart emails. For PGP/MIME, the whole MIME structure including all body parts is encrypted into a single part of content type *multipart/encrypted*.

### 2.2    S/MIME

In the late 1990s, S/MIME was specified as an Internet standard for email encryption and digital signatures based on X.509 public key certificates and a PKI. Besides having a more centralized trust model than OpenPGP, both standards have a lot in common. S/MIME and OpenPGP are both hybrid cryptosystems, consisting of a symmetric cipher such as AES and an asymmetric cipher like RSA. S/MIME encrypts the whole MIME structure into a single body part of content type *application/pkcs7-mime*. It is supported natively by various mail clients and used in business environments and organizations, such as universities.

### 2.3    MIME Email

Historically, RFC822 email was limited to ASCII messages. This did not fit the needs of users to send other file formats such as binary data. Therefore, in 1992 Multipurpose Internet Mail Extensions (MIME) were born, enabling emails that consist of multiple parts of various content types. An example HTML email with inline images, additional text parts, and a PDF attachment is given in Figure 2.

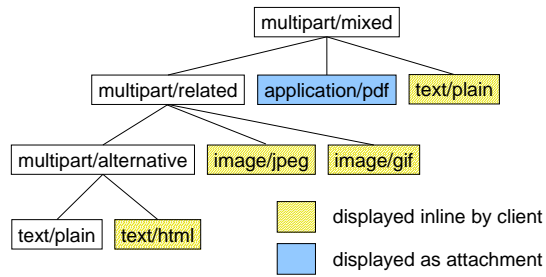

Fig. 2: Exemplary MIME tree of a multipart email.

In the context of end-to-end encryption, the flexibility of multipart mails can be dangerous. Neither OpenPGP nor the S/MIME standard cover the edge-case of partially encrypted messages: e.g., ciphertexts can be wrapped as a sub-part within the MIME tree, which is the foundation of our attacks on encryption.

### 2.4   HTML Email

HTML in emails was introduced by Netscape in 1995 to format messages, e.g., to provide bold or colored text. It competed with the *text/enriched* MIME type as defined in RFC1563 and Microsoft's proprietary Rich Text Format (RTF). HTML email was eventually adopted by the general public, despite opposition by tech enthusiasts (as expressed, e.g., in the ASCII ribbon campaign). Today most mail clients support HTML emails by default.[5] However, until today, there is no standard that defines which HTML elements should be enabled in email. For example, some email clients even execute script tags within emails (see [8]).

## 3   Related Work

In 2000 Katz, Schneier, and Jallad [6, 7] presented chosen-ciphertext attacks against OpenPGP and S/MIME, in which they make use of the malleability feature of CFB and CBC mode to modify encrypted messages resulting in 'garbage' plaintext. A victim replying to the garbled plaintext unwittingly acts as a decryption oracle, allowing the receiver to reconstruct the original plaintext. Heiderich et al. [5] showed that this attack is possible even without ciphertext masking. Recently, Poddebniak et al. [8] demonstrated that the malleability of CFB/CBC can be used to modify encrypted emails such that their plaintext is automatically exfiltrated to the attacker when opened in a vulnerable email client, using HTML and other backchannels. They, furthermore, showed that some email clients concatenate encrypted and unencrypted MIME parts, allowing an attacker to leak the plaintext of OpenPGP and S/MIME encrypted messages by loading them as the resource of a remote URL. Message takeover attacks for signed emails have been discussed by Davis [3] in 2001. He showed that a signed message "Let's break up" from Bob to Eve can simply be re-send by Eve to scare Alice (Bob's new girlfriend). Furthermore, Davis demonstrated that signatures can simply be removed in many scenarios and the message can be re-signed by the attacker. In 2017, Ribeiro [10] showed that the displayed content of signed HTML emails can be changed subsequently if the mail client fetches external CSS stylesheets.

## 4   Attacker Model

Attacks based on decryption oracles require the attacker to *somehow* have obtained PGP or S/MIME encrypted emails. In practice, this could be achieved via an untrustworthy or compromised SMTP or IMAP server, via a third party component such as cloud-based antivirus solutions scanning transiting emails, or via a compromised mailbox (e.g., based on weak passwords or XSS on the webmail service). While this is a strong attacker model, the only reason to use end-to-end encryption at all is that an untrusted communication channel is presumed.

---

[5] According to an email marketing statistics and metrics study conducted by Juniper Research, 97% of all email clients used in 2007 supported HTML messages.

After having obtained ciphertext messages, the attacker, Eve, can re-send them in her own name to one of the original communication parties, Alice or Bob. Note that both can act as a decryption oracle because emails are usually encrypted with the public key of both, the sender and the receiver, as both parties want to be able to decrypt it later. Eve can perform additional changes to the encrypted messages such as wrapping them within a multipart mail. In addition, Eve may apply social engineering to lure the victim – Alice or Bob – into replying to her (benign-looking) message. Note that this is a weak requirement as it is a basic function of email to reply to communication partners, even previously unknown ones. It is clear that the security of a cryptographic protocol should not be dependent on the assumption that no communication is made. Signing oracle-based attacks only require the victim to reply to a benign-looking email.

## 5   Decryption Oracles

Replying to a decrypted email and quoting the original message can leak the plaintext to a third party in case the *From:* or *Reply-To:* header had been replaced with the attacker's email address. Such message takeover attacks under a new identity are well-known (see [6, 7]). However, they can often be detected based on the message content. It is generally assumed that trained users should get suspicious and discover the deception instead of replying to 'out of context' messages. In this paper we show how to hide the original plaintext and instead show a meaningful message, asking the user to reply and, therefore, leak the (hidden) plaintext. We do this by abusing the MIME standard in combination with HTML email. Encrypted messages can themselves be a sub-part within a MIME tree which may include further non-encrypted parts. Even though there are hardly meaningful use cases for such 'partially encrypted' emails, they are a valid feature. This allows an attacker to integrate captured ciphertext messages into a MIME tree under her control and re-send this new email to the victim (i.e., the original sender or receiver). A MIME tree containing an attacker-controlled message, as well as S/MIME and OpenPGP encrypted parts, is given in Figure 3.

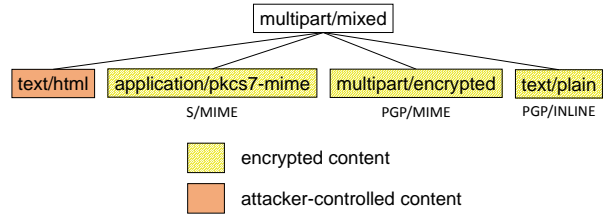

Fig. 3: MIME tree of a partially encrypted email.

**Plaintext Merged with Attacker's Text.** If a client receives a multipart email, it decrypts the ciphertext parts and afterwards merges all ASCII and HTML parts into a single document which is quoted upon replying.[6] This implementation approach of the MIME standard can be considered dangerous: Eve can prepend her own message, followed by a lot of newlines, to the captured ciphertext part. If Alice replies without scrolling down in her reply quoting the text, she unintentionally acts as a decryption oracle and leaks the plaintext. Other obfuscation techniques include hiding the ciphertext somewhere between the attacker's message parts: Emails, especially forwarded mails, can contain a long conversation history and top-posting without reading the whole conversation history is common user behavior. A user replying to a 'mixed content' conversation can thereby leak the plaintext of encrypted messages wrapped within the attacker-controlled text.

**Plaintext Hidden Using HTML and CSS.** In the context of HTML email, mixed content attacks are more serious than in ASCII emails. An attacker who can inject her own HTML/CSS code into the same document where the plaintext is displayed can completely hide it, e.g., by wrapping it within an iframe. An example email is given in Figure 4. The result for Apple Mail is shown in Figure 5.

```
1   From: eve@evil.com
2   To: johnny@good.com
3   Content-Type: multipart/mixed; boundary="BOUNDARY"
4
5   --BOUNDARY
6   Content-Type: text/html
7
8   <b>Hello Johnny,</b>
9   I'm interested in your work. Could you explain to me how...
10  <iframe height="1" frameborder="0">
11  --BOUNDARY
12  Content-Type: application/pkcs7-mime; smime-type=enveloped-data
13  Content-Transfer-Encoding: base64
14
15  [... ciphertext ...]
16  --BOUNDARY--
```

(a) Attacker-prepared multipart email received by victim's mail client.

```
1
2
3
4   <b>Hello Johnny,</b>
5   I'm interested in your work. Could you...
6   <iframe height="1" frameborder="0">
7   Secret message, for Johnny's eye only...
```

```
1   Dear Eve, ...
2
3   On 01/05/19 08:27, Eve wrote:
4   > <b>Hello Johnny,</b>
5   > I'm interested in your work. Could you...
6   <iframe height="1" frameborder="0">
7   Secret message, for Johnny's eye only...
```

(b) HTML code after decryption.                (c) HTML code in reply message.

Fig. 4: Email structure to hide S/MIME ciphertext in an invisible *iframe*. After decryption the plaintext will be included as 'covert content' in the quoted reply.

---

[6] There are alternative ways to handle multipart messages. The email client "The Bat!" shows a new tab for each body part, while Outlook only displays the very first part. However, a majority of the evaluated clients follows the described approach.

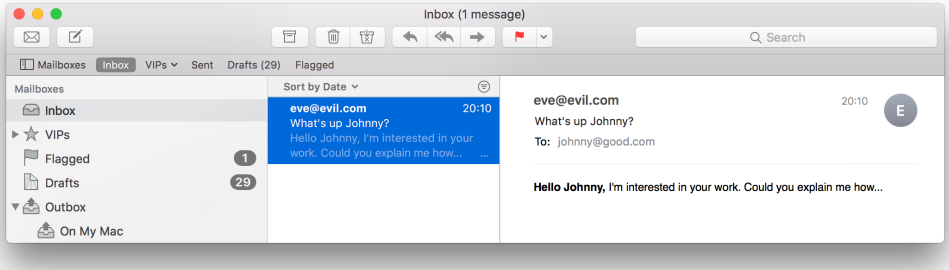

(a) Johnny receives a benign-looking email from Eve.

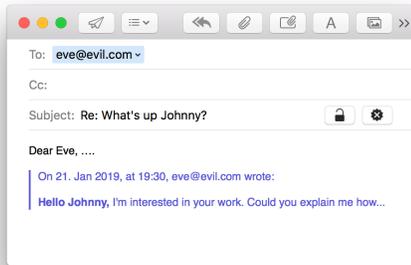

(b) Johnny replies to Eve.

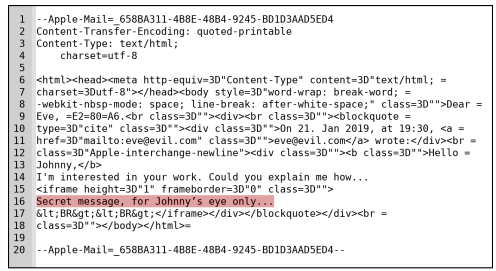

(c) Eve obtains the plaintext.

Fig. 5: Covert content attack using Apple Mail as S/MIME decryption oracle.

Note that a closing `</iframe>` tag is not required. However, it could easily be added by placing another attacker-controlled *text/html* part at the end of the message. Iframes are just one way to hide the original plaintext. Other options include wrapping it into HTML comments or other elements such as `<audio>` or `<canvas>` which do not display the content between opening and closing tags – while it is still kept when replying to the email. Other, more advanced, techniques to hide the plaintext using CSS properties are shown for attacks on signatures in section 6. A comprehensive list of CSS blinding options is given in Table 1.

**Breaking Mixed-Content Isolation with References.** In cases where multiple MIME parts are not automatically concatenated by the client, this behavior can be enforced by creating a *multipart/related* email structure referencing the ciphertext via *cid:* URI schemes (see RFC2392). Such Content-ID resource locators are typically used to embed and display inline images within HTML emails. They are generally seen as more compatible than referencing remote images which are blocked in most email clients for privacy reasons. In the example email given in Figure 6, the attacker's *text/html* part includes the ciphertext as an 'image'. Because the resulting plaintext is not a valid image file, it cannot be displayed by the client. However, the decrypted inline 'image' is included in reply emails, therefore leaking the plaintext. A resulting screenshot of the wrapped PGP/MIME message being opened in Thunderbird is given in Figure 10 in the appendix. The attacker is not limited to images; the plaintext can also be referenced as the content of an *iframe*, *object*, *embed*, and other elements.

```
1   From: eve@evil.com
2   To: johnny@good.com
3   Content-Type: multipart/related; boundary="BOUNDARY"
4
5   --BOUNDARY
6   Content-Type: text/html
7
8   What's up Johnny?
9   
10  <style>fieldset ,br{display:none}</style>
11
12  --BOUNDARY
13  Content-ID: <target>
14  Content-Type: multipart/encrypted; protocol="application/pgp-encrypted"; boundary="PGPMIME"
15
16  --PGPMIME
17  Content-Type: application/pgp-encrypted
18
19  Version: 1
20  --PGPMIME
21  Content-Type: application/octet-stream; name="encrypted.asc"
22  Content-Disposition: inline; filename="encrypted.asc"
23
24  -----BEGIN PGP MESSAGE-----
25  [... ciphertext ...]
26  -----END PGP MESSAGE-----
27  --PGPMIME--
28  --BOUNDARY--
```

Fig. 6: Email structure to hide PGP/MIME ciphertext in a referenced 'image'.

Note that the attack does not require a 'partially encrypted' email because Eve can also encrypt her malicious parts with the victim's public PGP key or S/MIME certificate. The attack is even successful if the victim replies to Eve with an encrypted email because Eve's public key is used for re-encryption. These attacks apply not only for single ciphertext messages in the middle part of a multipart email, but hundreds of encrypted emails can be hidden as sub-parts and their plaintext can be leaked with a single reply.[7] Furthermore, the attack does not require an active MitM, but rather, the obtained ciphertext could be years-old. For example, a nation-state actor could have captured a target user's encrypted emails over years and later decides to expose them by sending a single benign-looking email which lures the user into replying. While the attacks use email to exfiltrate the plaintext, their scope is not limited to exfiltrating decrypted emails. The attacks also work with non-email ciphertexts such as PGP encrypted files. Covert content attacks are independent of the applied encryption scheme, even though email clients and crypto plugins may handle multipart messages differently, depending on whether S/MIME and OpenPGP is used. While the attacks require user interaction, they do not require any 'unusual' behavior, but instead normal usage of email as a communication medium. They also do not require complex cryptographic attacks like the CBC gadgets discussed in [8].

---

[7] At some point, the SMTP server may enforce a resource limit, e.g., 25 MB for Gmail.

## 6    Signing Oracles

Digital signatures should guarantee integrity, authenticity, and non-repudiation of messages. To give an example, Johnny could be a commander-in-chief who takes information security seriously. All his emails are digitally signed, making it hard to impersonate him in order to send forged statements or instructions. The goal of our attacker Eve is to start false-flag warfare. Therefore, she needs to obtain a digitally signed 'declaration of war' which she can forward to the armed forces. Every time Johnny replies to a message he already acts – to a certain extent – as a signing oracle when quoting the original text. For example, consider the following message from Eve to Johnny:

```
1   I hereby declare war.
```

Johnny replies with a signed message, thereby quoting the original text:

```
1   Sorry Eve, You can't do that.
2
3   On 01/05/19 09:42, Eve wrote:
4   > I hereby declare war.
```

In the reply, commander Johnny unintentionally signed Eve's quoted text. Certainly, given the message context and the quote prefix (>...) it is clear that declaring war is not his intention. However, Eve can try to hide her malicious content using *CSS blinding options* while a benign text message, such as *"What's up Johnny?"*, is added to be shown. Similarly, the benign text can be hidden while showing the malicious content, based on *CSS conditional rules* which are satisfied only for a third party. If Johnny replies to such a specially-crafted HTML/CSS email, he signs arbitrary covert content along with visible content. This signed message can then be forwarded by Eve to a third party (e.g., the armed forces) where it displays the previously hidden malicious content *"I hereby declare war"*, while hiding the benign content. A schematic illustration of such covert content attacks on email signatures is given in Figure 7.

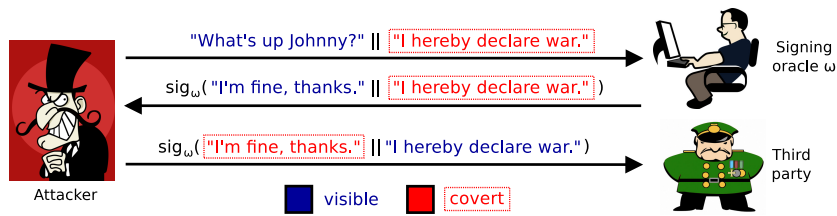

Fig. 7: Covert content attacks against email signatures.

A simple HTML email containing conditional CSS code to display different content based on the device's screen resolution is given in Figure 8. It can be used to obtain a signed email from a mobile device, where a benign message is shown. The reply message instead displays a (signed) declaration of war when shown on a desktop mail client. A screenshot of the attack using iOS Mail as a signing oracle and the resulting signed email shown in Thunderbird is given in Figure 9.

```
1   From: eve@evil.com
2   To: johnny@good.com
3   Content-Type: text/html
4
5   <style>
6   /* hide malicious content on mobile devices */
7   @media (max-device-width: 834px) {
8   .covert {visibility: hidden;}
9   }
10  /* but show on desktop/large-screen devices */
11  @media (min-device-width: 835px) {
12  * {visibility: hidden;}
13  .covert {visibility: visible !important; position: absolute; top: 8px; left: 8px;}
14  }
15  </style>
16
17  What's up Johnny?
18  <div class="covert" style="visibility: hidden">I hereby declare war.</div>
```

(a) Attacker-prepared HTML/CSS email sent to Johnny.

```
1   What's up Johnny?
```

(b) Content seen by Johnny on his mobile email client.

```
1   I'm fine, thanks.
2
3   On 01/05/19 09:53, Eve wrote:
4   > What's up Johnny?
```

(c) Content seen by Johnny when replying to the message.

```
1   I hereby declare war.
```

(d) Signed content seen by a third party on a desktop client.

Fig. 8: Malicious HTML/CSS email to obtain a signed 'declaration of war'.

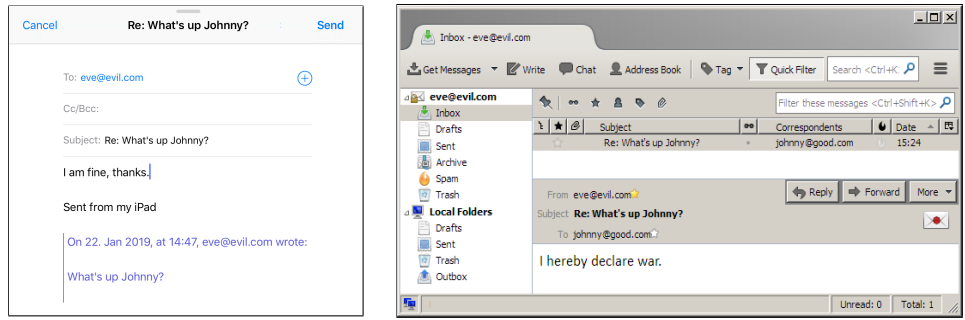

(a) Johnny replies to Eve's email.

(b) Eve obtains a signed reply email for arbitrary text.

Fig. 9: Covert content attack abusing iOS Mail as S/MIME signing oracle.

In the given example, email clients with a screen width of less than 835px (e.g., a mobile phone or tablet) show a different text than desktop mail clients based on the `@media` conditional rule. If the email client includes this conditional CSS in the reply message it can be misused as a signing oracle, therefore allowing the attacker to obtain signed messaged for arbitrary (displayed) content.

**Conditional Rules.** The W3C specifies CSS conditional rules [12] like `@media`, which allow different formatting based on conditions such as screen width or orientation. For example, a different text can be shown whether a mobile phone is held in portrait or landscape mode, or whether the document is displayed on a screen or printed out. Besides media queries, we can show different text in different email clients using the `@support` conditional rule, which applies formatting based on CSS feature support in the client. For example, an email can be shown in red if two property-value pairs are supported:

```
1   @supports (property1: value1) and (property2: value2) {* {color:red}}
```

We assembled a list of over 1,000 CSS property-value pairs to fingerprint the features supported by clients. This allows us to selectively enable certain CSS code for every client that interprets the `@support` rule. A further conditional rule introduced by Mozilla is `@document`. It allows CSS code to be executed based on the document location. In the context of email clients, this even allows us to show different text for each user because the location contains an *imap://* URI scheme with the email address. For example, to apply a red color solely for the emails of *general@good.com* the following CSS code can be used:

```
1   @-moz-document url-prefix("imap://general@good.com") {* {color:red}}
```

In case CSS conditional rules are not supported, email clients may support their own proprietary conditional statements. For example, Outlook interprets HTML and CSS code within `<!--[if mso]>...<![endif]-->`, while other clients will ignore it. A listing of other conditional features is given in Figure 11 in the appendix.

**Blinding Options.** We identified seven CSS properties which can be used for covert content attacks, as shown in Table 1. However, this list is unlikely to be complete because CSS is very complex and offers more possibilities to hide text.

| property | show | hide |
|---|---|---|
| display: | initial; | none; |
| visibility: | visible; | hidden; |
| opacity: | 1; | 0; |
| clip-path: | initial; | polygon(0px 0px, 0px 0px, 0px 0px, 0px 0px); |
| position: | static; | absolute; top: -9999px; left: -9999px; |
| color: | initial; | transparent; |
| font-size: | initial; | 0; |

Table 1: CSS properties to hide text.

The proposed attacks allow an attacker to obtain valid signatures for arbitrary content to be displayed. This can be used to trick a third party, which relies on the authenticity and integrity of signed messages, to perform certain actions (such as starting a war). A forensic analysis can reveal the deception, but then it may already be too late (i.e., war is already declared). Note that the covert content attacks to obtain signatures do not require any MIME wrapping, but rather depend on HTML emails, and on support for (internal) CSS styles.

## 7  Evaluation

To evaluate the proposed attacks, we selected 19 widely-used email clients with OpenPGP support and 22 clients supporting S/MIME from a comprehensive list of over 50 email clients assembled from public software directories for all major platforms (Windows, Linux, macOS, Android, iOS, and Web). Email clients were excluded if they were not updated for several years, or if the cost to obtain them would be prohibitive (e.g., appliances). All clients were tested in the default settings with an additional PGP or S/MIME plugin installed where required. The results from the tested clients regarding covert content attacks, (i.e., tricking a user into acting as an oracle for decryption or signing) are shown in Table 2.

| | | *Support* | | *Decryption* | | *Signatures* | |
|---|---|---|---|---|---|---|---|
| | | **S/MIME** | **PGP** | **S/MIME** | **PGP** | **S/MIME** | **PGP** |
| Windows | Thunderbird (52.5.2) | native | Enigmail | ● | ● | ● | ● |
| | Outlook 2016 (16.0.4266) | native | GpgOL | ○ | ○ | ◐ | ● |
| | Win. 10 Mail (17.8730) | native | – | ○ | – | ◐ | – |
| | Win. Live Mail (16.4.3528) | native | – | ○ | – | ● | – |
| | The Bat! (8.2.0) | native | GnuPG | ○ | ○ | ○ | ○ |
| | Postbox (5.0.20) | native | Enigmail | ● | ● | ● | ● |
| | eM Client (7.1.31849.0) | native | native | ● | ○ | ◐ | ◐ |
| Linux | KMail (5.2.3) | native | GPGME | ◐ | ◐ | ○ | ○ |
| | Evolution (3.22.6) | native | GnuPG | ◐ | ◐ | ◐ | ◐ |
| | Trojitá (0.7-278) | native | GPGME | ◐ | ◐ | ● | ◐ |
| | Claws (3.14.1) | plugin | GPG plugin | ◐ | ◐ | ○ | ○ |
| | Mutt (1.7.2) | native | GPGME | ◐ | ◐ | ○ | ○ |
| macOS | Apple Mail (11.2) | native | GPGTools | ● | ● | ◐ | ◐ |
| | MailMate (1.10) | native | GPGTools | ● | ● | ● | ● |
| | Airmail (3.5.3) | plugin | GPG-PGP | ● | ● | ● | ● |
| iOS | Mail App (11.2.2) | native | – | ● | – | ● | – |
| Android | K-9 Mail (5.403) | – | OpenKeychain | – | ● | – | ● |
| | R2Mail2 (2.30) | native | native | ○ | ● | ◐ | ◐ |
| | MailDroid (4.81) | Flipdog | Flipdog | ○ | ● | ● | ● |
| | Nine (4.1.3a) | native | – | ○ | – | ● | – |
| Web | Exchange/OWA (15.1.1034) | plugin | – | ○ | – | ● | – |
| | Roundcube (1.3.4) | plugin | Enigma | – | ◐ | ◐ | ◐ |
| | Horde/IMP (6.2.21) | native | GnuPG | ○ | ○ | ◐ | ◐ |
| | Mailpile (1.0.0rc2) | – | GnuPG | – | ○ | – | ○ |

decryption oracles { ● Plaintext can be completely hidden  ● Covert rules are kept in reply message } signing oracles
◐ Plaintext merged with attacker-text  ◐ Covert rules only for received message
– Cryptosystem not available  ○ No vulnerabilities found

Table 2: Evaluation of covert content attacks on email encryption and signatures

All tested email clients quote the original message when replying, which is the precondition for our attacks. Of the overall tested 24 clients, 20 display HTML emails in the default settings without any additional user interaction, but only 16 clients reply with HTML formatted content. While only five clients download external CSS style sheets by default, all HTML capable clients support internal and/or inline CSS, and at least one blinding option to hide text. All but two HTML capable clients support conditional rules or other features to conditionally show or hide text. Full details on HTML and CSS support for the various tested email clients are given in Table 3 in the appendix.

### 7.1  Decryption Oracles

All email clients, excluding Microsoft products and "The Bat!", merge multiple ASCII text or HTML parts into a single document when replying, making them potentially vulnerable to covert content attacks. However, not all clients decrypt ciphertext sub-parts within the MIME tree, thereby disabling the attack. From discussions with application developers, we learned that this was initially *not* meant as a security precaution. Instead, the case of partially encrypted messages was simply not considered in the implementation of S/MIME or the PGP plugin. As a consequence, clients that are more feature complete, have higher compatibility, and require a larger implementation effort are more likely to be misused as decryption oracles. We consider clients as vulnerable if the plaintext of encrypted messages can either be completely hidden, or if it is concatenated with attacker-controlled text.

For seven clients, including popular applications such as Apple Mail or Thunderbird, we could completely hide the ciphertext within a multipart mail using HTML/CSS and show arbitrary content instead. A user replying to such a benign-looking email unknowingly leaks the plaintext of up to hundreds of encrypted emails at once. For another six vulnerable clients, HTML formatted replies are deactivated in the default settings or not supported at all. In such cases, our attacks are limited because the decrypted message cannot be completely hidden. However, it can be appended to the attacker's text, separated by a lot of newlines, or wrapped somewhere within the conversation history. All affected clients, except R2Mail2, show consistent behavior, independently of whether S/MIME or OpenPGP is used as encryption scheme.

### 7.2  Signing Oracles

We classify clients as vulnerable not only if they can act as a signing oracle, but also if they show different text for signed messages based on conditional CSS. Both vulnerabilities are required for the attack, but they do not need to exist in the same client. In fact, because the targeted users (e.g., *Johnny* and *General*) in each of these cases are different, they are likely to use different clients.

Ten clients, including popular applications such as Thunderbird, K-9 Mail, the iOS Mail App, and Outlook Web Application (OWA), the GUI for Microsoft Exchange, keep the original `<style>` element in replies, allowing an attacker to misuse them as signing oracles.[8] Of the remaining clients, six convert internal CSS style information into inline styles when replying and eight clients reply to HTML emails with ASCII text in the default settings. Once a signed email with conditional CSS has been obtained, it can be used to trick 18 of the 20 clients displaying HTML in the default settings (all but Mailpile and "The Bat!") as well as the HTML-to-text converter used by Horde/IMP into selectively showing/hiding certain text. We could observe the same behavior for all email clients, independent of the applied encryption scheme.

---

[8] It must be noted that for two clients, MailMate and Airmail, some additional effort was required to bypass filters which would otherwise strip internal CSS styles.

## 8    Countermeasures

Building a secure encryption protocol on top of email is very challenging. There are many pitfalls and edge-cases to be considered. In this section, we provide best practices to counter the attacks previously described. These practices should be of help to guide implementations of OpenPGP or S/MIME capable clients.

### 8.1    Decryption Oracles

**All-or-Nothing Encryption.** Partially encrypted messages can be considered harmful. Therefore, email clients *must not* decrypt emails unless they contain a single encrypted part (i.e., the root node in the MIME tree). This can be standardized and enforced for S/MIME and PGP/MIME. For PGP/Inline however, the only way to send a multipart message is to separately encrypt each part. Unfortunately, every PGP/MIME message can be interpreted in the context of a PGP/Inline message (i.e., a downgrade attack). Hence, email clients supporting PGP/Inline *must* enforce a strict separation between multiple body parts, for example, by opening each part in a separate window or tab. When replying to multipart messages, only the very first body part may be quoted and, therefore, included in the reply to prevent unintended leakage of covert plaintext content.

**Accepting ASCII Text Only.** Active content such as HTML within emails is dangerous. Disabling HTML prevents most attacks described in this work. Unfortunately, this does not meet today's usage of email. HTML email has become the norm and in ten of the tested email clients – for example, in Apple Mail and iOS Mail – there is not even an option to disable HTML for incoming emails. It must be additionally noted that modern email clients also display *text/plain* emails within an HTML widget component. One major problem is that no definition for 'HTML email' exists. Developing a standard describing a 'safe' subset of HTML which can be used in emails to allow basic formatting, but forbid potentially harmful features, would be a step in the right direction.

**Enforcing Digital Signatures.** In theory, signed emails offer protection against covert content attacks. If Bob received an email originating from Eve, but one message part was signed by Alice, he may get suspicious and not reply to Eve. In practice, email clients miserably fail when it comes to verifying signatures for multipart messages. Our tests show that most email clients either do not show a signature at all for partially signed messages, or show the first available signature in the MIME tree – which can originate from Eve because she can simply re-sign the message. Even in cases where the client explicitly shows inline information regarding which part is signed, we managed to hide the signature information itself using CSS. Moreover, S/MIME signatures can be stripped by targeted modifications of the CBC-ciphertext as shown by Strenzke [11]. Nevertheless, digital signatures – if done right – can enhance message authenticity and integrity. For example, a company could set up a policy to discard all incoming messages if they do not contain exactly one single sign-then-encrypt message part, including signed email headers which can be enforced using extensions such as Memory Hole for OpenPGP [4] or Secure Header Fields for S/MIME [1].

It is important to note that the described countermeasures must be implemented by *all* involved parties. Usually, a user has no control over the security precautions taken by his communication partners. In the context of email end-to-end encryption, this is problematic because both the sender and the receiver can act as a decryption oracle for captured ciphertext. Even if Bob discarded partially encrypted messages and disabled HTML, Alice may still be vulnerable.

### 8.2   Signing Oracles

**Dropping CSS Support.** Conditional CSS makes it easy for an attacker to hide certain text within a signed message while showing different text. Ideally, clients would ignore CSS in received emails. However, this is an unrealistic scenario given today's usage of email, especially in a business context, where it is expected that emails can have any sort of formatting – technically implemented with CSS. Sanitizing conditional CSS rules and properties which can be used to hide content is feasible, but it may be insufficient as web technologies are constantly evolving. Nevertheless, it is important to display digitally signed content equally to all viewers. The S/MIME and OpenPGP standards, which are from a time-period where messages were ASCII text, fail to address this and should be extended.

**Only ASCII Text in Replies.** It should not harm the user experience if mail clients converted quoted messages into ASCII text when replying to an email. Eight of the tested clients (e.g., Roundcube) are actually doing this. Thus, we recommend that security-focused mail clients should adopt this behavior. They *must not* sign any quoted HTML/CSS input from the original message, so that they cannot be misused as signing oracles.

## 9   Conclusion

Email is complex. The MIME standard and HTML, as supported by modern email clients, provide a high level of flexibility and allow arbitrary wrapping, nesting, and hiding of encrypted or to-be-signed content. This complexity and the conjoined attack surface are not dealt with in the security considerations of the OpenPGP and S/MIME standards, which primarily focus on cryptographic algorithms and their parameters such as key sizes. However, relying on the security of cryptographic primitives, such as AES or ECDH, is not enough for secure email end-to-end encryption and signatures. The developers of email clients have to handle a plethora of critical edge-cases – without being able to consult any published best practices. Our work aims to close this research gap. We reveal implementation pitfalls in the "no man's land" between cryptography and email, as used today, and give guidance and best practices in order to improve the security of S/MIME and OpenPGP capable email clients.

## Acknowledgements

The authors thank Juraj Somorovsky for his valuable feedback and insightful discussions. Jens Müller was supported by the research training group 'Human Centered System Security' sponsored by the state of North-Rhine Westfalia. In addition, this work was supported by the German Research Foundation (DFG) within the framework of the Excellence Strategy of the Federal Government and the States – EXC 2092 CASA.
Note this is a draft version. The final version of this work will be published at the 17th International Conference on Applied Cryptography and Network Security.

# A    Screenshots of Decryption Oracles

## A.1    Plaintext Hidden in a Referenced Inline 'Image'.

Figure 10 depicts a covert content attack against Thunderbird/Enigmail based on the example email given in Figure 6. The ciphertext is hidden in an embedded 'image' file, referenced from the attacker's part via a *cid:* URI scheme. The OpenPGP plugin – Enigmail – detects the 'image' as PGP/MIME content and decrypts it. The decrypted 'image' is then Base64 encoded by Thunderbird and included in the reply message, therefore leaking the plaintext.

(a) Johnny receives a benign-looking email with an embedded invisible 'image' which contains PGP/MIME ciphertext.

(b) Johnny replies to Eve, thereby unknowingly leaking the plaintext within the invisible inline image.

(c) Eve obtains the reply email, including the Base64 encoded 'image' which contains the plaintext.

(d) Eve decodes the Base64 encoded data, resulting in the original plaintext MIME message.

Fig. 10: Convert content attack using Thunderbird as PGP decryption oracle.

## B  HTML/CSS Email Support

| | HTML | | CSS styles | | | blinding options | | | | | | | conditional rules | | | |
|---|---|---|---|---|---|---|---|---|---|---|---|---|---|---|---|---|
| | view | reply | external | internal | inline | display | visibility | opacity | clip-path | position | color | font-size | @media | @supports | @document | other |
| **Windows** Thunderbird | ● | ● | ◐ | ● | ● | ● | ● | ● | ● | ● | ● | ● | ● | ○ | ○ | ○ |
| Outlook 2016 | ● | ● | ◐ | ● | ● | ● | ○ | ○ | ○ | ○ | ● | ○ | ○ | ○ | ○ | ○ |
| Win. 10 Mail | ● | ● | ◐ | ● | ● | ● | ● | ○ | ○ | ○ | ● | ○ | ○ | ○ | ○ | ○ |
| W. Live Mail | ● | ● | ● | ● | ● | ● | ● | ○ | ○ | ● | ● | ○ | ○ | ○ | ○ | ○ |
| The Bat! | ● | ● | ◐ | ● | ● | ○ | ● | ● | ○ | ● | ● | ○ | ○ | ○ | ○ | ○ |
| Postbox | ● | ● | ◐ | ● | ● | ● | ● | ● | ● | ● | ● | ● | ● | ○ | ○ | ○ |
| eM Client | ● | ● | ● | ● | ● | ● | ● | ● | ● | ● | ● | ● | ○ | ○ | ○ | ○ |
| **Linux** KMail | ◐ | ● | ○ | ◐ | ◐ | ◐ | ◐ | ◐ | ○ | ◐ | ◐ | ◐ | ◐ | ○ | ○ | ○ |
| Evolution | ● | ◐ | ◐ | ● | ● | ● | ● | ● | ● | ● | ● | ● | ● | ○ | ○ | ○ |
| Trojitá | ● | ○ | ◐ | ● | ● | ● | ● | ● | ● | ● | ● | ● | ● | ○ | ○ | ○ |
| Claws | ○ | ○ | ○ | ○ | ○ | ○ | ○ | ○ | ○ | ○ | ○ | ○ | ○ | ○ | ○ | ○ |
| Mutt | ○ | ○ | ○ | ○ | ○ | ○ | ○ | ○ | ○ | ○ | ○ | ○ | ○ | ○ | ○ | ○ |
| **Mac** Apple Mail | ● | ● | ◐ | ● | ● | ● | ● | ● | ● | ● | ● | ● | ○ | ○ | ● | ● |
| MailMate | ● | ● | ● | ● | ● | ● | ● | ○ | ● | ● | ● | ○ | ○ | ○ | ● | ● |
| Airmail | ● | ● | ◐ | ● | ● | ● | ● | ● | ● | ● | ● | ● | ● | ○ | ● | ● |
| **iOS** Mail App | ● | ● | ● | ● | ● | ● | ● | ● | ● | ● | ● | ● | ● | ○ | ● | ● |
| **Android** K-9 Mail | ● | ● | ◐ | ● | ● | ● | ● | ● | ● | ● | ● | ● | ● | ○ | ○ | ○ |
| R2Mail2 | ● | ○ | ◐ | ● | ● | ● | ● | ● | ● | ● | ● | ● | ● | ○ | ○ | ○ |
| MailDroid | ● | ● | ◐ | ● | ● | ● | ● | ● | ● | ● | ● | ● | ● | ○ | ○ | ○ |
| Nine | ● | ● | ○ | ○ | ○ | ○ | ○ | ○ | ○ | ○ | ○ | ○ | ○ | ○ | ○ | ○ |
| **Web** Exchange/OWA | ● | ● | ○ | ○ | ○ | ● | ● | ● | ○ | ● | ● | ○ | ○ | ○ | ○ | ● |
| Roundcube | ● | ● | ◐ | ● | ● | ● | ● | ● | ● | ● | ● | ● | ● | ● | ● | ● |
| Horde/IMP | ◐ | ◐ | ● | ◐ | ◐ | ◐ | ◐ | ◐ | ◐ | ◐ | ◐ | ◐ | ◐ | ◐ | ● | ● |
| Mailpile | ● | ○ | ◐ | ● | ○ | ● | ● | ● | ● | ● | ● | ● | ○ | ○ | ○ | ● |

○ Not supported by client    ● Supported in default settings    ◐ Supported in non-default settings

Table 3: HTML and CSS support in various email clients.

## C  Other Conditional Features

```
1   <html><head>
2   <!--[if IE]><style>.wlm {color: red;}</style><![endif]-->      <!-- Windows Live Mail -->
3   <!--[if mso]><style>.ol {color: red;}</style><![endif]-->      <!-- Outlook / W10Mail -->
4   <style>
5   .ExternalClass .owa, [owa] .owa {color: red;}                  /* Exchange (OWA) */
6   .moz-text-html .tb {color: red;}                               /* Thunderbird */
7   </style>
8   </head>
9   <body>
10  <div class="wlm"> RED text only in Windows Live Mail </div>
11  <div class="ol"> RED text only in Outlook / W10Mail </div>
12  <div class="owa"> RED text only in Exchange (OWA)     </div>
13  <div class="tb"> RED text only in Thunderbird         </div>
14  </body></html>
```

Fig. 11: Proprietary features and CSS to target only certain clients.